\documentclass[letterpaper]{article}

\usepackage{emulateapj}
\usepackage{onecolfloat}
\usepackage{apjfonts}
\usepackage{amsmath}
\usepackage{natbib}
\usepackage{graphicx}

\begin{document}

\submitted{Accepted by the Astrophysical Journal}

\twocolumn[
\title{X-Ray Observations of XTE~J1550--564 During the Decay of
the 2000 Outburst -- I. {\em Chandra} and {\em RXTE} Energy Spectra}

\author{John A. Tomsick}
\affil{Center for Astrophysics and Space 
Sciences, University of California, San Diego, MS 0424, La Jolla, CA 92093 
(e-mail: jtomsick@ucsd.edu)}

\author{St\'ephane Corbel}
\affil{Universit\'e Paris VII and Service d'Astrophysique, CEA Saclay, 
91191 Gif sur Yvette, France 
(e-mail:corbel@discovery.saclay.cea.fr)}

\author{Philip Kaaret}
\affil{Harvard-Smithsonian Center for Astrophysics, 60 Garden Street, 
Cambridge, MA 02138 (e-mail: pkaaret@cfa.harvard.edu)}

\begin{abstract}

We report on the evolution of the X-ray energy spectrum for the black 
hole candidate (BHC) X-ray transient XTE~J1550--564 during the decay 
of the 2000 outburst.  The {\em Rossi X-ray Timing Explorer (RXTE)} and 
{\em Chandra} observations span nearly five orders of magnitude in 
luminosity.  The {\em RXTE} spectra are dominated by a power-law 
component and a comparatively weak soft component was also detected 
for the first two observations.  The source made a transition to
the canonical hard state near a luminosity of 
$9\times 10^{36}$~erg~s$^{-1}$ over several observations as evidenced
by a drop in the flux of the soft component in the {\em RXTE} 
energy band and a hardening of the power-law component to a photon
index near 1.6.  The power-law did not exhibit this behavior for 
the previous XTE~J1550--564 outburst.  For some observations, we 
detect a high energy cutoff and find that the cutoff is more significant 
and at lower energy during the transition than in the hard state.
The cutoff in the hard state is at higher energy than has been seen for 
most previous accreting BHCs.  The {\em Chandra} spectrum provides 
evidence for spectral evolution after the hard state transition.  It is 
well, but not uniquely, described by a power-law with a photon index of 
$2.30^{+0.41}_{-0.48}$ (90\% confidence) and interstellar absorption.  
Advection-dominated accretion flow models predict gradual spectral 
softening as the luminosity drops, but our observations do not allow us 
to determine if the spectral evolution is gradual or sudden.  The lowest 
luminosity we measure for XTE~J1550--564 with {\em Chandra} is 
$5\times 10^{32}$~erg~s$^{-1}$ (0.5-7~keV, for a distance of 4~kpc).
Although this is probably not the true quiescent luminosity, it represents
a useful upper limit on this quantity.

\end{abstract}


] 

\section{Introduction}

The energy spectra of accreting black hole candidates (BHC), while complex, 
are dominated by two emission components:  a soft thermal component peaking 
below 10~keV and a hard component that extends to hundreds of keV \citep{grove98}.  
For most BHC X-ray transients, both components are present in the X-ray band when 
the source is bright during the outburst, and the source is said to be in the soft 
state or the very high state depending on the details of the spectral and timing 
properties (see van der Klis 1995\nocite{vdk95} for a review of spectral states).  
There is strong evidence that the soft component is optically thick blackbody 
emission from the accretion disk.  As the X-ray luminosity drops, significant 
changes in the spectral and timing properties are observed as sources enter the 
hard state.  The soft component becomes undetectable at energies above 1~keV, 
and the spectrum is dominated by a hard power-law or breaking power-law 
component.  An increase in the strength of the timing noise is also characteristic 
of the transition to the hard state.  It is likely that the high energy emission is due 
to inverse Comptonization of soft photons by energetic electrons, but there is 
uncertainty about the system geometry, the electron energy distribution and
the mechanism for transferring energy to the electrons.  One possibility is that a 
quasi-spherical, optically thin region forms in the inner portion of the accretion 
disk as predicted by advection-dominated accretion flow (ADAF) models 
\citep{ngm97}.  Another possibility that implies a different site for hard X-ray 
production is that the hard X-ray emission is due to magnetic flares above the 
disk \citep{grv79}.  The source behavior during the decay from the hard state into 
quiescence is not well-established.  Here, we present a study of the evolution
of the energy spectrum for the BHC X-ray transient XTE~J1550--564 spanning
nearly five orders of magnitude in luminosity during outburst decay.

XTE~J1550--564 was first detected by the {\em Rossi X-ray Timing Explorer 
(RXTE)} All-Sky Monitor in 1998 September \citep{smith98}.  It was 
identified as a probable black hole system based on its X-ray spectral 
\citep{sobczak99} and timing properties.  The source shows strong aperiodic 
variability including low (0.08-18~Hz) and high (100-285~Hz) frequency 
quasi-periodic oscillations 
\citep{cui99,remillard99,homan99,remillard01,homan01,miller01b}.
The optical \citep{obj98} and radio \citep{campbell98} counterparts were 
identified, and recent optical observations confirm that the system contains 
a black hole \citep{orosz01}.  A superluminal ejection was observed in the
radio \citep{hannikainen01}, establishing that the source is a microquasar 
similar to GRO~J1655--40 and GRS~1915+105.  XTE~J1550--564 became active in 
X-rays again in 2000 April \citep{smith00}, making it one of the few BHC X-ray 
transients for which multiple outbursts have been observed.  X-ray observations 
during the 2000 outburst were made by us and also by another group 
\citep{miller01a}.  A radio study of the 2000 outburst \citep{corbel01} 
showed evidence for a compact jet when the 
source was in the hard state and that this jet is quenched during the 
intermediate/very high state (see Homan et al.~2001\nocite{homan01} for a 
discussion of spectral states in XTE~J1550--564).  In 2001 January, the source 
flared again and was detected in the X-ray \citep{tomsick01}, optical \citep{jbt01} 
and radio, but this flare did not develop into a full outburst.

In this paper, we focus on the properties of the energy spectrum
during the decay of the 2000 outburst.  The {\em RXTE} and {\em Chandra} 
observations used for this paper include a study of a transition to 
the hard state and spectral evolution at low luminosities.  
A study of the timing properties using the {\em RXTE} data is presented in
a companion paper (Kalemci et al.~2001\nocite{kalemci01}, hereafter paper II).

\section{Observations and Light Curve}

\begin{figure}[t]
\plotone{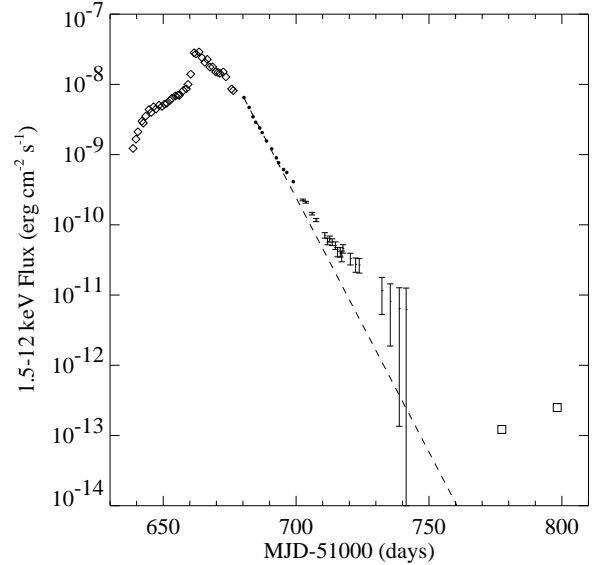}
\caption{The X-ray light curve for the 2000 outburst from XTE~J1550--564, 
including data from the {\em RXTE}/ASM (diamonds), {\em RXTE} pointed 
observations (filled circles and error bars) and {\em Chandra} 
observations (squares).  The dashed line is an exponential decay with
an e-folding time of 6.0 days.\label{fig:lightcurve}}
\end{figure}

During the XTE~J1550--564 outburst, we obtained two {\em Chandra} 
observations of the BHC at low flux and several observations with 
{\em RXTE} during the decay of the outburst.  The {\em RXTE} observations 
were made under a program to study BHC X-ray transients during outburst 
decay, and the {\em Chandra} observations were granted from Director's 
Discretionary Time motivated by our {\em RXTE} observations.  
Figure~\ref{fig:lightcurve} shows the 1.5-12~keV light curve for the outburst, 
including data from the {\em RXTE} All-Sky Monitor and our pointed 
observations.  Table~\ref{tab:rxte_obs} provides information about the 
observations that we used for spectral analysis.  

When observing sources near the Galactic plane with {\em RXTE}, one
must be aware of contributions from the Galactic ridge emission \citep{vm98}
and the possibility of source confusion.  
XTE~J1550--564 is $1^{\circ}.8$ from the Galactic plane, and the Galactic 
ridge emission is significant for our fainter observations.  To estimate the 
level of Galactic ridge emission, we used public {\em RXTE} observations of the 
XTE~J1550--564 position that were made when the source was in quiescence.  
The observations were made on 1999 August 13 and consist mostly of scans 
used to study the nearby supernova remnant G326.3--1.8.  Although the 
observations only include 128 seconds on the XTE~J1550--564 position, this 
is sufficient to determine the X-ray flux.  Assuming an absorbed Raymond-Smith 
plus power-law model with the parameter values found by \cite{vm98} for the 
central ridge but with the overall normalization left as a free parameter, the 
Galactic ridge flux is $1.68\times 10^{-11}$~erg~cm$^{-2}$~s$^{-1}$ (1.5-12~keV) 
at the XTE~J1550--564 position.

Source confusion is also an issue for our {\em RXTE} observations.  
The transient pulsar XTE~J1543--568 \citep{marshall00}, which is 
$0^{\circ}.966$ from the XTE~J1550--564 pointing position \citep{kaptein01}, 
was in outburst during our observations.  We analyzed the ASM data for 
XTE~J1543--568 to determine the flux this source contributed during our 
observations.  For ten 15 day intervals during the XTE~J1550--564 outburst, 
the ASM count rate for the pulsar is between 0.15 and 0.90~cps.  At the 
maximum count rate and given the 3\% response that is expected for the
angular separation,  we estimate that XTE~J1543--568 contributes a flux of
$1.08\times 10^{-11}$~erg~cm$^{-2}$~s$^{-1}$ (1.5-12~keV) during our
observations.  The level of emission observed during our last two {\em RXTE} 
observations is consistent with the emission coming from the Galactic ridge 
and XTE~J1543--568 with little or no contribution from XTE~J1550--564.  The 
emission properties observed for our last several observations confirm that 
we are seeing significant contributions from the Galactic ridge and 
XTE~J1543--568.  We find a strong iron line at 6.7~keV for our last several 
observations, which is a property of the Galactic ridge emission 
\citep{vm98}.  Also, we detect pulsations at 27 seconds, which is the
known pulse period for XTE~J1543--568 \citep{marshall00}.  

For the light curve shown in Figure~\ref{fig:lightcurve} we assume that 
the flux contribution to the {\em RXTE} observations from the Galactic ridge 
emission and XTE~J1543--568 is between 
$1.68\times 10^{-11}$~erg~cm$^{-2}$~s$^{-1}$ and the value we find 
for our last observation, 
$2.94\times 10^{-11}$~erg~cm$^{-2}$~s$^{-1}$ (1.5-12~keV).  
For the first 13 {\em RXTE} observations, the level of contamination
is not significant.  These observations are shown as filled circles in
Figure~\ref{fig:lightcurve}, and we used only these observations
for the spectral analysis described below.  Error bars are shown
for the other {\em RXTE} observations.  Fitting an exponential
to the first 13 {\em RXTE} observations gives 6.0 days for the
e-folding time for the decay, which is a very fast decay compared to 
other BHC X-ray transients \citep{csl97,tk00}.  A deviation from the 
exponential decay first occurs on MJD 51695 during observation 11.  
It is interesting that a flare is seen in the near-IR light curves for 
XTE~J1550--564 that peaks at this time \citep{jain01} and that the 
radio observation indicating the presence of a compact jet was made 
soon after on MJD 51697 \citep{corbel01}.  A fractionally larger deviation 
from the exponential decay in the X-ray light curve occurs between MJD 51710 
and 51730.  Generally, the X-ray light curve for XTE~J1550--564 is similar to 
those of other BHC X-ray transients, which typically show reflares, glitches and 
secondary maxima \citep{csl97}.  We note that the ASM data for 
XTE~J1543--568 indicates that it is unlikely that features in the 
XTE~J1550--564 light curve are due to flux variations from the pulsar.

The {\em Chandra} observations were made on MJD 51777.405 
(2000 August 21) and MJD 51798.245 (2000 September 11) using the 
ACIS (Advanced CCD Imaging Spectrometer).  We used one of the 
back-illuminated ACIS chips (S3) to obtain the best low energy 
response.  For the first observation, a 1.5-12~keV flux of 
$1.2\times 10^{-13}$~erg~cm$^{-2}$~s$^{-1}$ is inferred (assuming 
an absorbed power-law model), while the flux was 
$2.5\times 10^{-13}$~erg~cm$^{-2}$~s$^{-1}$ for the second 
observation.  The fact that the flux increased by more than a 
factor of 2 between the observations indicates that the main part 
of the decay had ended by the time the {\em Chandra} observations 
were made.  This is consistent with the optical behavior of the 
source.  XTE~J1550--564 reached optical quiescence by MJD 51750 
and remained at its quiescent level during both of our {\em Chandra} 
observations \citep{jain01}.

\begin{figure}
\plotone{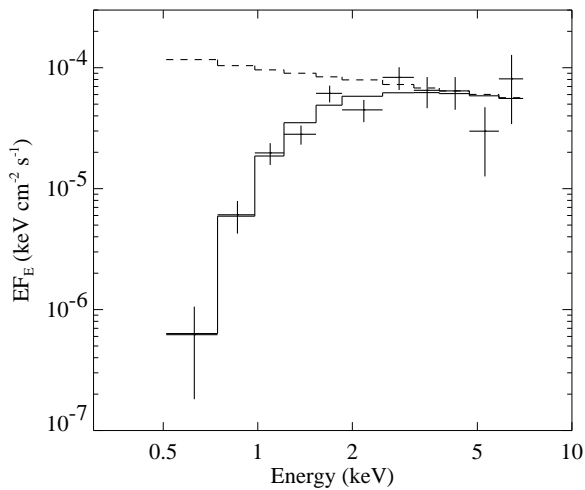}
\caption{The {\em Chandra}/ACIS energy spectrum for the two {\em Chandra}
observations combined.  The solid line is a power-law with interstellar
absorption, and the dashed line is a power-law with $\Gamma = 2.3$ 
without absorption.\label{fig:chandra_spectrum}}
\end{figure}

\section{{\em Chandra} Analysis and Results}

We analyzed the data from the {\em Chandra} observations using the 
CIAO v2.1 and XSPEC v11.0 software packages.  XTE~J1550--564 is 
clearly detected in both observations at R.A. = 15h 50m 58s.65, 
Decl. = --56$^{\circ}$~28$^{\prime}$~35.$^{\prime\prime}$2 (equinox 2000.0) 
with an uncertainty of 1$^{\prime\prime}$, which is consistent with the 
optical and radio positions previously reported \citep{jain99,corbel01}.  
We extracted source spectra for each observation from a circular region 
with a radius of $5^{\prime\prime}$ and background spectra from an annulus 
with an inner radius of $5^{\prime\prime}$ and an outer radius of 
$18^{\prime\prime}$ centered on the source.  The outer radius is 
constrained by the presence of a nearby source.  For the first observation, 
there are brief time segments where excess background is observed.  
Removing these causes a drop in the exposure time from 5099~seconds 
to 4985~seconds.  Periods of high background are not seen in the 
second observation allowing us to use the entire 4572~seconds.  In the 
energy band used for spectral analysis (0.5-7~keV), 71 and 111 counts 
were collected in the source region for the first and second observations, 
respectively.  We determined that the expected background level in the 
source extraction region in the 0.5-7~keV energy band is 1.2 and 1.3 counts 
for the first and second observations, respectively, indicating that background 
subtraction is unnecessary.  The CIAO routine ``psextract'' was used to 
extract the spectra and create response matrices.  It is important to note 
that the response matrices created by this routine do not account for the
energy dependence of the High Resolution Mirror Array (HRMA) point 
spread function (PSF).  This leads to underestimating the source flux, 
especially at higher energies where the PSF is broader.  To correct the 
response matrix for this effect, we used the HRMA calibration data from 
Table~4.2 of the {\em Chandra} Proposer's Guide ($10^{\prime\prime}$ 
diameter case).  The data was interpolated to obtain a value for the 
encircled energy fraction for each energy bin in the ACIS response matrix, 
and the correction was applied by multiplying the effective areas used for 
the response matrix by these fractions.  For our $5^{\prime\prime}$ extraction 
radius, the correction causes an increase of the 0.5-7~keV energy flux (and 
thus the quoted luminosities) by 5\%.  It should be noted that this correction 
becomes significantly larger if small extraction radii are used.

We began by fitting each spectrum with a power-law model, which is 
commonly used for fitting BHC X-ray transients at low flux levels 
(e.g., Asai et al.~1998\nocite{asai98}).  We included interstellar absorption
and left the column density as a free parameter.  
Although the flux is different by a factor of 2 between observations, 
the spectral parameters obtained are not significantly different, and 
we carried out further spectral analysis after combining the data for 
the two observations.  The increase in flux for the second observation 
is rather surprising, and we made light curves for the two observations 
with four and eight time bins to determine if the flux increase was 
caused by a short time scale phenomenon (e.g., a flare).  We found no 
statistically significant variability in either observation.

We rebinned the spectrum for the two observations combined as shown 
in Figure~\ref{fig:chandra_spectrum} and fitted the spectrum using 
$\chi^{2}$-minimization first with a power-law model and then with a 
simple blackbody (both with absorption).  The fit is somewhat
better for the power-law model ($\chi^{2}/\nu = 9.5/8$) than for
the blackbody model ($\chi^{2}/\nu = 12.8/8$).  In addition, a 
blackbody temperature of 0.7~keV is obtained, which is considerably 
higher than typical values seen for quiescent neutron star transients 
\citep{asai98}.  The distance for XTE~J1550--564 has been estimated 
at between 2.5~kpc \citep{sanchez99} and 6~kpc \citep{sobczak99}.
For the flux observed during the {\em Chandra} observations, 
a blackbody temperature of 0.7~keV implies a blackbody radius
of between 0.06~km and 0.15~km for distances of 2.5~kpc and
6~kpc, respectively, indicating that it is unlikely that a blackbody 
interpretation could be physically meaningful.  These findings 
are consistent with the identification of XTE~J1550--564 as a black 
hole rather than a neutron star system, and, in the following, we 
focus on the power-law model.

\begin{figure}[t]
\plotone{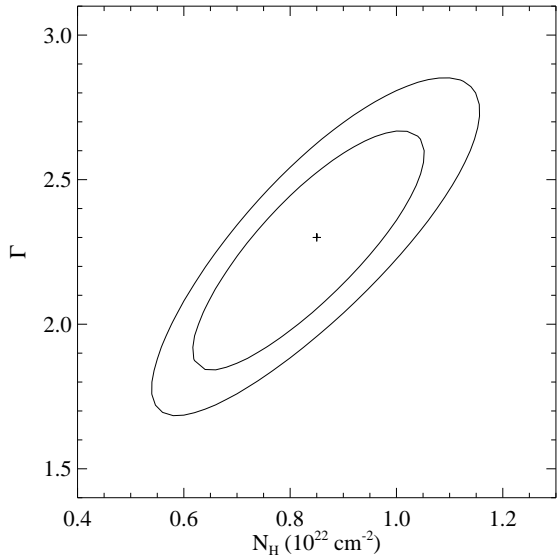}
\caption{Error contours for the column density ($N_{\rm H}$) and
the power-law index ($\Gamma$) derived from the {\em Chandra}
spectrum.  The plus marks the location of the best fit values
and the 68\% ($\Delta\chi^{2} = 2.30$) and 90\% 
($\Delta\chi^{2} = 4.61$) confidence contours are shown.
\label{fig:contour}}
\end{figure}

We refitted the spectrum for the two observations combined with a 
power-law model with interstellar absorption using the Cash statistic 
\citep{cash79}.  This maximum likelihood method allows for the 
determination of errors and confidence intervals for parameters and 
does not require that the data be rebinned in energy.  This is 
desirable since rebinning data necessarily removes spectral information.  
We obtain values of $\Gamma = 2.30^{+0.41}_{-0.48}$ for the power-law 
photon index and $N_{\rm H} = (8.5^{+2.2}_{-2.4})\times 10^{21}$~cm$^{-2}$ 
for the column density (90\% confidence errors are given in both cases).  
We note that the correction for the energy dependence of the PSF described 
above only changes $\Gamma$ by 2\%, which is insignificant compared to
the statistical errors.  Figure~\ref{fig:contour} shows the 68\% and 
90\% error contours for the two parameters.  The column density is 
consistent with $9\times 10^{21}$~cm$^{-2}$, which is the value for 
the Galactic column density along the line of sight \citep{jain99}, 
but it is only marginally consistent with the value inferred from 
the optical work of \cite{sanchez99}.  From optical spectra of 
XTE~J1550--564 taken during outburst, they find that
E(B-V) = $0.70\pm 0.10$, which corresponds to a column density
of $(3.9\pm 0.6)\times 10^{21}$~cm$^{-2}$ \citep{ps95}.  
The value of $N_{\rm H}$ measured by {\em Chandra} could
indicate that the distance to the source is somewhat greater 
than 2.5~kpc, which is the distance estimate obtained from 
the measured value of E(B-V).  The significance of the value we 
find for $\Gamma$ is discussed below.  

\section{{\em RXTE} Spectral Analysis and Results}

We produced Proportional Counter Array (PCA) and High Energy X-ray Timing 
Experiment (HEXTE) energy spectra for each observation (see Bradt, 
Rothschild \& Swank 1993\nocite{brs93} for instrument descriptions).  
We used the PCA in the 3-25~keV energy band and HEXTE in the 18-150~keV 
energy band.  For the PCA, we used standard mode data, consisting of 129-bin 
spectra with 16~second time resolution, included the photons from all three 
anode layers and estimated the background using the ``Sky-VLE'' model.  
For each observation, data was combined from the Proportional Counter Units 
(PCUs) 1-4 that were turned on.  For most of the observations, at least one 
of the PCUs was turned off.  Damage to PCU 0 prevented us from using data 
from this detector.  We produced HEXTE energy spectra using 
standard mode data, consisting of 64-bin spectra with 16~second time resolution.  
We used the 1997 March response matrices and applied the necessary dead-time 
correction \citep{rothschild98}.  For the spectral fits, the normalization was 
left free between HEXTE and the PCA.  The HEXTE background subtraction is 
performed by rocking on and off source.  Each cluster has two background fields, 
and we checked the background subtraction by comparing the count rates for 
the two fields.  

For the PCA, we used the 2001 February response matrices and tested 
them by combining spectra from all the available Crab observations from 
2000 May 14 to 2001 Jan 28 where PCUs 1-4 were turned on, resulting in a 
Crab spectrum with an exposure time of 17,584~seconds.  We fitted the 
spectrum with a model consisting of two power-law components, representing 
the contributions from the nebula and the pulsar, with interstellar 
absorption.  We fixed the column density to 
$3.2\times 10^{21}$~cm$^{-2}$ \citep{massaro00} and the photon index for 
the pulsar component to 1.8 \citep{knight82}.  We restricted the bandpass
to 3-25~keV since very large residuals (near or above the 5\% level) are 
present outside this range.  Even when only the 3-25~keV range is used, we 
obtain a reduced chi-squared well above 1.0 due to the small statistical errors.
The Crab residuals are larger below 8~keV than above this energy, and 
including 0.8\% systematic errors from 3-8~keV and 0.4\% systematic
errors from 8-25~keV leads to a drop in the reduced chi-squared to 1.0.
We added these systematic errors to the XTE~J1550--564 spectra to account 
for uncertainties in the PCA response.  

\begin{figure}[t]
\plotone{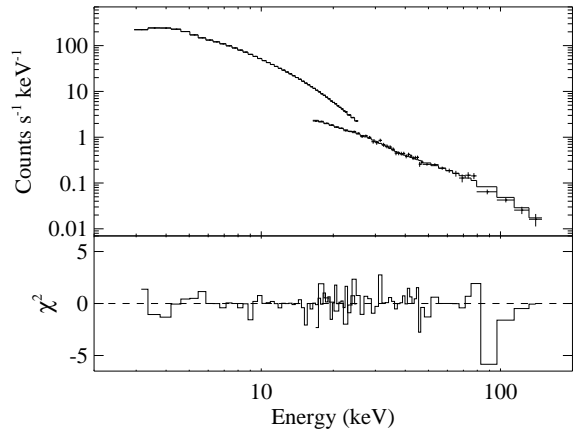}
\caption{The {\em RXTE} energy spectrum for observation 1, including
the PCA from 3-25~keV and HEXTE from 18-150~keV.  The bottom panel
shows the residuals in the form of the contribution to the value
of $\chi^{2}$ for each energy bin.
\label{fig:rxte_spectrum}}
\end{figure}

We determined the spectral model to use by fitting the PCA plus HEXTE energy 
spectrum for observation 1.  A power-law alone (with interstellar absorption) 
provides a very poor fit with $\chi^{2}/\nu = 1118/90$.  Large positive residuals 
are seen near 6-7~keV, and a broad minimum centered near 10~keV is present in 
the residuals.  Models with cutoffs at high energies, such as the Comptonization
model of \cite{titarchuk94}, give even worse fits ($\chi^{2}/\nu = 1649/88$)
since the spectrum extends to high energies without a clear cutoff.
We obtain a significant improvement in the fit to $\chi^{2}/\nu = 135/87$ by 
adding a smeared iron edge \citep{ebisawa94} to the power-law.  The smeared 
edge has been used by other authors to fit XTE~J1550--546 spectra 
\citep{sobczak00} and also for other BHC transients during outburst decay 
\citep{tk00}.  Even with the smeared edge, large residuals are present below 
4~keV, suggesting the presence of a soft component.  The addition of a 
disk-blackbody model \citep{makishima86} with a temperature near 0.5~keV 
improves the fit significantly to $\chi^{2}/\nu = 71/85$.  Finally, the
addition of a gaussian iron line improves the fit to $\chi^{2}/\nu = 59/82$, 
which is significant at the 99.8\% level based on an F-test.  Although we left all 
the model parameters free for these fits, the data is not adequate to provide 
good constraints on the column density, the width of the smeared edge, the line 
energy or the line width, and, for subsequent fits, we fixed these parameters as 
follows.  We used a value of $9\times 10^{21}$~cm$^{-2}$ for the column density, 
which is the Galactic column density along the line of sight to XTE~J1550--564 
and is also consistent with the value found using {\em Chandra}.  Following 
\cite{sobczak00}, we fixed the width of the smeared edge to 7~keV and the width 
($\sigma$) of the iron line to 0.5~keV.  Also, we fixed the line energy to 6.5~keV, 
which is consistent with the K$\alpha$ line for moderately ionized iron.  We note
that all these values fall within the 68\% confidence error regions for the
parameters when they are left free, and that we obtain $\chi^{2}/\nu = 63/86$ 
when the parameters are fixed.  Figure~\ref{fig:rxte_spectrum} shows the 
observation 1 spectrum and the residuals for a model consisting of a power-law, 
a disk-blackbody, a smeared edge and an iron line with the parameters fixed as 
specified above.

We used the same model to fit the other XTE~J1550--564 observations.  The
model provides a good description of the spectrum in each case with reduced
chi-squared values between 0.56 and 1.26 for between 54 and 87 degrees of 
freedom.  We only include results for observations 1-13 because the 
contamination from the Galactic ridge and the pulsar XTE~J1543--568 provides 
a significant source of uncertainty in parameter estimation for later observations.  
F-tests indicate that the disk-blackbody is only detected at extremely high 
significance ($1-2\times 10^{-15}$) for the first observation.  It is marginally 
detected at 98\% confidence for the second observation, and is not statistically
significant for the other observations.  Inner disk temperatures ($kT_{in}$)
of 0.70 and 0.62~keV are obtained for observations 1 and 2, respectively.
Although the statistical error on the temperature measurements is 0.05~keV, 
there is also a systematic uncertainty related to the fact that $kT_{in}$ 
depends on the column density.  We estimate that the total error on the 
temperature measurements for observations 1 and 2 is near 0.1~keV.  
The disk-blackbody normalizations\footnote{The disk-blackbody normalization
is $N = \cos i~(R_{in}/d_{10})^{2}$ where $i$ is the binary inclination, 
$R_{in}$ is the accretion disk inner radius in units of km and $d_{10}$ is 
the source distance in units of 10~kpc.} are $533^{+202}_{-147}$ and 
$355^{+1145}_{-210}$ for observations 1 and 2, respectively.  As for the 
temperature, the 68\% confidence statistical errors reported here underestimate 
the true normalization uncertainties.  Although the disk-blackbody component 
is not detected at high significance for observations 3-13, we obtain values 
of $kT_{in}$ near 0.35~keV for observations 3 and 4 with this component
in the model.  In fitting the spectra for observations 3-13, we included a 
disk-blackbody component with the temperature fixed to 0.35~keV.  At 
such a low temperature, this component only contributes to the lowest two 
or three PCA energy bins and does not significantly change the values for 
the other fit parameters.  For observations 3-5, we calculate 90\% confidence 
upper limits on $N$ between 9000 and 15000, and, for observations 6-13, the 
upper limits are between 1000 and 2100.

The smeared edge is detected at greater than 99\% confidence for all 
observations except for 11 and 13.  The edge energies are consistent
with neutral or moderately ionized iron and the optical depth of the edge
\citep{ebisawa94} drops over time from 0.9 for observation 1 to about 0.4 
for observations 8, 9, 10 and 12.  However, we emphasize that the smeared 
edge model is phenomenological, and physical significance should not be 
attached to its parameters.  The feature itself is clearly present in most 
of the spectra and may be related to Compton reflection \citep{lw88}.  
For most of the observations, the iron line equivalent widths are between
15 and 75~eV, and the highest equivalent width of 110~eV occurs for
observation 13.  The detection significance for the line is greater than 
99\% for six of the observations.

One of our main results is the evolution of the power-law index during the
decay of the outburst.  Table~\ref{tab:spec} shows the photon index for the 
13 observations using the PCA alone, HEXTE alone and for both 
instruments combined.  In all cases, the photon index is near 2.0 for 
observation 1, but gradually drops to about 1.6 during the next few observations.  
We fitted the spectra for the two instruments separately to obtain an estimate
of the systematic error due either to the model we are using for the low energy 
portion of the spectrum or due to uncertainty in the cross-calibration between the
PCA and HEXTE.  Although the differences between the PCA and HEXTE
are only statistically significant for a few observations, it is clear that there is
a systematic effect since the index measured by the PCA is softer than the
index measured by HEXTE in 11 of the 13 observations.  We estimate the 
systematic error in the measurement of the photon index by calculating the 
weighted average of the differences between the PCA and HEXTE.
This gives a value of 0.03, which we adopt as the systematic error on the
measurement of the power-law index. 

\begin{figure}[t]
\plotone{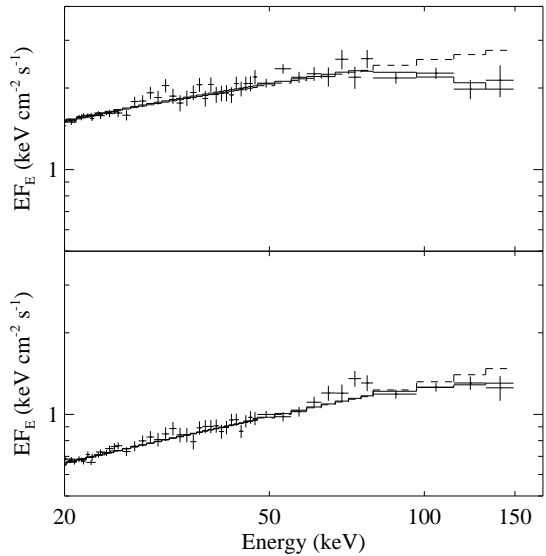}
\caption{The high energy portions of the {\em RXTE} spectra for 
observation 3 (top panel) and observations 5-12 combined (bottom 
panel).  In both cases, the dashed line shows the fit to the 
spectrum using a power-law with no high energy cutoff and the
solid line shows the fit using a power-law with a high energy
cutoff.  It is clear that a more significant cutoff is present
for observation 3, during the state transition, than for 
observations 5-12, when the source was in the hard state.
\label{fig:highenergy}}
\end{figure}

Although the spectral model described above provides satisfactory fits
to the energy spectra, for a few of the observations, the spectra appear 
to break in the HEXTE band.  To determine if the breaks are statistically
significant, we added a high energy cutoff to our model and refit the 
PCA/HEXTE energy spectra.  We used the XSPEC model ``highecut'' 
which has two free parameters and has been used previously
by several authors to model spectral breaks (e.g., Grove et al.~1998).  
Including this model modifies the power-law for energies above $E_{cut}$ 
producing an exponential cutoff with an e-folding energy $E_{fold}$ 
but does not alter the spectrum for energies below $E_{cut}$.  
For observations 2, 3 and 4, F-tests show that the cutoff is significant at 
99.8\% confidence or greater, but the cutoff is less significant for the other 
observations.  The cutoff is only detected at extremely high significance 
($1-2.6\times 10^{-8}$) for observation 3.  Table~\ref{tab:spec} includes the 
measurement of the photon index when the break is included for observations 
2, 3 and 4.  The change in photon index is not significant for any of the 
observations.  We combined the data from observations 5-12, which all have the 
same value for $\Gamma$ within errors, and refitted the PCA/HEXTE spectrum with 
and without a cutoff.  Adding a cutoff provides a marginally significant (98\%
confidence) improvement, indicating that the break is weak in this case if it is present.
The best fit values for $E_{cut}$ and $E_{fold}$ are given in Table~\ref{tab:cutoff} 
for observations 2, 3, 4 and 5-12.  Figure~\ref{fig:highenergy} shows the high 
energy portion of the spectrum for observation 3, where the most significant 
cutoff is detected, and for observations 5-12.  

\section{Discussion}

\subsection{Transition to the Hard State}

\begin{figure}[b]
\plotone{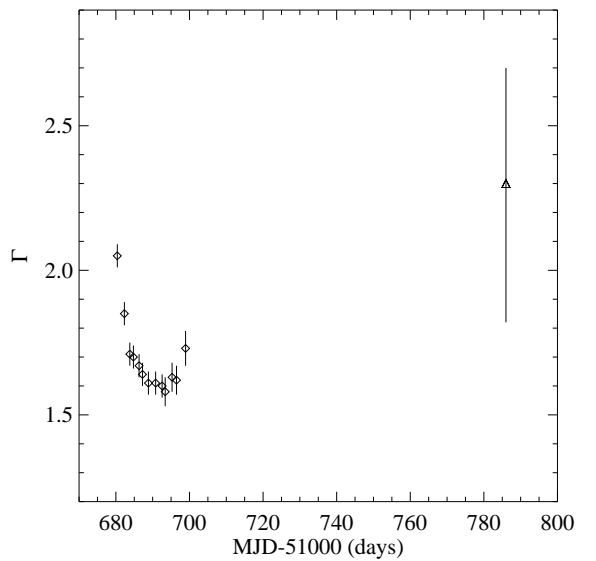}
\caption{The evolution of the power-law index ($\Gamma$) for the
pointed {\em RXTE} observations (diamonds) and the combination
of the two {\em Chandra} observations (triangle).  The {\em RXTE}
values come from the last column of Table~\ref{tab:spec} and the
error bars shown include the systematic error discussed in the
text.  For the {\em Chandra} point, a 90\% ($\Delta\chi^{2} = 2.71$)
confidence error bar is shown.\label{fig:gamma}}
\end{figure}

The spectral evolution we observe during the {\em RXTE} observations
shows that XTE~J1550--564 was in the process of making a transition to the 
hard state as our observations began.  Figure~\ref{fig:gamma} shows the
evolution of the power-law photon index ($\Gamma$) for the {\em RXTE} 
and {\em Chandra} observations.  The drop in $\Gamma$ during the first
few {\em RXTE} observations along with the timing properties reported in 
paper II indicate that the source made a transition to the hard state.  In 
addition, the drop in temperature and eventual non-detection of the soft 
component is common when such transitions occur 
(see Tomsick \& Kaaret 2000\nocite{tk00} and references therein).  Although 
our observations do not provide information about the state of the source
before the transition, \cite{corbel01} indicate that XTE~J1550--564 was in the 
intermediate/very high state before the transition since relatively strong hard 
X-ray emission was observed by BATSE (Burst and Transient Source Experiment) 
on {\em CGRO (Compton Gamma-Ray Observatory)} throughout the outburst.  
Observations of XTE~J1550--564 made during the rise of the 1998-1999 outburst 
also indicate that the source made a transition from the hard state to the 
intermediate/very high state \citep{cui99,wd01}.

XTE~J1550--564 is one of the few BHC X-ray transients that has had multiple
outbursts, allowing us to compare the source behavior during the 2000 decay
to that of the previous outburst.  For the 2000 outburst, the transition to 
the hard state occurred when the source was still relatively bright.  
The flux for the second {\em RXTE} observation, which is probably near the 
midpoint of the transition, is $5\times 10^{-9}$~erg~cm$^{-2}$~s$^{-1}$ 
(3-25~keV).  The previous outburst was observed with {\em RXTE} to a flux
level over an order of magnitude lower \citep{sobczak00}.  Although the 
behavior of the soft component shows some similarities between the two outbursts, 
the evolution of the power-law component was completely different.  In contrast
to the 2000 decay, $\Gamma$ remained at between 2.2 and 2.6 down to a 
2-20~keV flux level of $5\times 10^{-10}$~erg~cm$^{-2}$~s$^{-1}$ for the
previous outburst.  The subsequent observations showed some hardening of the
power-law component, but the energy spectrum did not show clear evidence for 
a transition to the hard state.  For both outbursts, a disk-blackbody component
with an inner disk temperature of 0.6-0.7~keV was present when the source 
flux was near $6\times 10^{-9}$~erg~cm$^{-2}$~s$^{-1}$, and, in both cases, 
a drop in temperature occurred as the source decayed.  As discussed by 
\cite{sobczak00}, the evolution of the soft component could indicate an 
increase in the inner radius of the accretion disk.  If this is the case, it is 
likely that the inner radius increased more rapidly with flux for the 2000 decay 
than for the previous decay since the soft component was detected down to very
low flux levels during the 1998-1999 decay, while we did not detect the soft 
component after observation 2.

We also used {\em RXTE} to monitor the BHC X-ray transient 4U~1630--47 
during the decay of its 1998 outburst \citep{tk00}.  The source made a transition 
to the hard state, and we observed 4U~1630--47 several times during the 
transition, making it interesting to compare its behavior to that of 
XTE~J1550--564.  For 4U~1630--47, $\Gamma$ hardened from 2.3 to 1.8 
over a period of 8~days and the soft component became undetectable during
the transition.  Although our observations did not include the beginning of 
the XTE~J1550--564 transition, $\Gamma$ changed from 2.05 to 1.67 in
3.4~days, suggesting that the transition time scale is similar for the two 
sources.  The estimated transition luminosities are similar for the two sources.  
Taking the second XTE~J1550--564 observation as the midpoint of the transition 
and assuming isotropic X-ray emission gives a transition luminosity of 
$9\times 10^{36}$~erg~s$^{-1}$ (3-25~keV, unabsorbed).  As mentioned above, 
distance estimates for XTE~J1550--564 range from 2.5~kpc to 6~kpc, and here we 
adopt a distance of 4~kpc.  For 4U~1630--47, we estimated a transition luminosity 
of $7\times 10^{36}$~erg~s$^{-1}$ (2.5-20~keV).

\subsection{Spectral Evolution after the Hard State Transition}

After the source made a transition to the hard state, Figure~\ref{fig:gamma}
shows that the power-law index remained very close to 1.6 until the last 
{\em RXTE} observation, where $\Gamma$ showed a marginally significant
increase to $1.73\pm 0.06$.  Although statistical significance of the
change is not high, spectral softening at low luminosities is predicted if 
an ADAF is present \citep{emn97} since the X-ray emission is dominated by 
inverse Comptonization from the thermal distribution of electrons in the 
ADAF region.  As the mass accretion rate drops, the predicted spectrum 
gradually softens since the optical depth of the ADAF region decreases, 
leading to a decrease in the Compton y-parameter.  The {\em Chandra}
spectrum provides evidence for futher softening at low luminosities, 
which is also a prediction of the ADAF model.  As shown in 
Figure~\ref{fig:gamma}, the lower limit on $\Gamma$ for the {\em Chandra}
observation is 1.82 based on the 90\% confidence error bars, representing
reasonably good evidence for spectral softening.  

Although there is evidence that spectral evolution occurred for 
XTE~J1550--564 between the {\em RXTE} observations and the {\em Chandra} 
observations, it is not clear if this evolution occurred gradually (as predicted
by ADAF models) or suddenly as the source flux decayed.  In fact, not many 
X-ray observations of BHC systems have been made in the luminosity gap 
between our last {\em RXTE} observation, where the 3-25~keV luminosity 
was $9.6\times 10^{35}$~erg~s$^{-1}$, and our {\em Chandra} observations, 
where the mean unabsorbed 0.5-7~keV luminosity was $6.7\times 10^{32}$~erg~s$^{-1}$,
and only a few of these observations have produced useful energy spectra.
{\em BeppoSAX (Satellite per Astronomia X)} observations of the BHC
GX~339--4 at $7\times 10^{33}$~erg~s$^{-1}$ indicate that it remains
relatively hard ($\Gamma = 1.64\pm 0.13$) at low luminosity
\citep{kong00}.  Similarly, an {\em ASCA (Advanced Satellite for Cosmology 
and Astrophysics)} observation of XTE~J1748--288 at a luminosity of 
$4\times 10^{35}$~erg~s$^{-1}$ (assuming a distance of 8.5~kpc) gave 
$\Gamma = 1.56^{+0.34}_{-0.31}$ \citep{kotani00}.  However, {\em Ginga} 
observations of GS~1124--68 show some evidence for spectral softening.  
A power-law index of $\Gamma = 1.84^{+0.03}_{-0.04}$ was measured at 
a source luminosity of $5\times 10^{34}$~erg~s$^{-1}$ (assuming a distance
of 2.5~kpc) even though $\Gamma = 1.5$ was obtained when the source
was brighter.  V404~Cyg is the one BHC X-ray transient with a quiescent 
luminosity higher than the XTE~J1550--564 luminosity during our {\em Chandra} 
observations \citep{garcia01}.  This source has been observed in quiescence 
at a luminosity near $10^{33}$~erg~s$^{-1}$ by both {\em ASCA} and 
{\em BeppoSAX}, and power-law indeces of $2.1^{+0.5}_{-0.3}$ \citep{nbm97}, 
$1.7^{+0.3}_{-0.2}$ \citep{asai98} and $1.9^{+0.6}_{-0.5}$ 
\citep{cps01} have been reported.  This provides some evidence for 
spectral softening at low luminosities since values of $\Gamma$
between 1.3 and 1.5 were observed for V404~Cyg during outburst
(Tanaka \& Lewin 1995\nocite{tl95}).

The results discussed in the previous paragraph appear to indicate that the
evolution of $\Gamma$ as the luminosity decreases is not consistent from
source-to-source.  Spectral softening is observed for GS~1124--68 and 
is probably observed for XTE~J1550--564 and V404~Cyg, while the observations
show that the spectra of GX~339--4 and XTE~J1748--288 remain hard down
to relatively low luminosities.  The latter type of behavior is difficult to explain 
with the ADAF model, but such behavior can be explained if the X-ray emission 
is due to magnetic flares above the disk \citep{grv79,dcf99}.  In fact, a magnetic 
flare model has been used to explain the spectral evolution of GX~339--4 \citep{dcf99}.
A sharp transition to a quiescent state would be expected if the emission in the hard 
state is due to magnetic flares.  This is because the MHD turbulence that is responsible 
for the magnetic flares \citep{bh91} is expected to turn off when the accretion disk 
temperature drops \citep{gm98}.  Other accretion physics may also be important at 
low luminosities that will alter the X-ray emission properties.  For low viscosity 
accretion flows, theory and simulations show that convection plays an important 
role \citep{nia00}.  In the X-ray band, convection-dominated accretion flows 
(CDAFs) produce most of their emission via thermal bremsstrahlung in contrast 
to the mechanisms at work for the ADAF and magnetic flare models.  Rather than 
being a pure power-law, CDAF models predict that a peak will be observed 
in the X-ray band \citep{bnq01}.  Although the XTE~J1550--564 
{\em Chandra} spectrum does not show evidence for such a peak, the 
statistical quality of the data does not allow us to rule this out either.

The radio observations of XTE~J1550--564 during the 2000 outburst
give evidence for the presence of a compact jet only after the
transition to the hard state \citep{corbel01}.  An inverted radio
spectrum was observed on MJD 51697.14 within a day of our
observation 12. Multi-wavelength observations suggest that the
synchrotron emission from the jet extends from the radio to at 
least the near-IR and optical range, indicating a very powerful 
compact jet \citep{corbel01}. None of the models discussed 
here (ADAF, CDAF and magnetic flares) include outflows required to 
produce such a jet. Further theoretical modeling, including outflows, 
would be of great interest to allow a comparison with the broad band 
(radio, IR, optical, and X-ray) behavior of the source.  This is 
currently under investigation for BHC systems in the hard state
(e.g., Markoff, Falcke \& Fender 2001\nocite{markoff01}).

\subsection{XTE~J1550--564 and X-Ray Transients at Low Luminosities}

X-ray observations of transient systems in quiescence suggest that, for a 
given mass accretion rate, black hole systems are significantly less luminous 
than neutron star systems \citep{garcia01}.  Since the ADAF model interprets 
this as being due to the presence of an event horizon for the black hole 
systems \citep{ngm97}, it is important to establish this trend using as many 
systems as possible, and it is interesting to compare the minimum luminosity 
we observe for XTE~J1550--564 to the quiescent luminosities of other 
transients.  During our first {\em Chandra} observation, the unabsorbed 
0.5-7~keV luminosity was between $2\times 10^{32}$~erg~s$^{-1}$ 
and $1.2\times 10^{33}$~erg~s$^{-1}$ based on the 2.5-6~kpc distance range.  
Although XTE~J1550--564 is an established black hole system, this range 
of luminosities is comparable to neutron star systems and is 25 to 150 
times brighter than the median quiescent black hole luminosity for detections
given in \cite{garcia01}.  While this could have important implications
for the question of whether the quiescent luminosities of black holes and
neutron stars are different, we believe that it is unlikely that the system
was in true quiescence during our {\em Chandra} observations.  The luminosity 
increase observed for our second {\em Chandra} observation and the 
mini-outburst that occurred about four months later both suggest 
the continuation of significant variability in the mass accretion rate.
Perhaps XTE~J1550--564 will prove to be similar to GRO~J1655--40, which
was detected at relatively high X-ray luminosity between two outbursts, 
but was later observed by {\em Chandra} with an order of magnitude lower 
luminosity.  Since our observations only provide an upper limit on the true 
quiescent luminosity, future X-ray observations of XTE~J1550--564 are necessary 
to determine this quantity.

The {\em Chandra} spectrum for XTE~J1550--564 provides an opportunity 
for comparison to the neutron star system Cen~X-4.  This comparison is 
especially useful because Cen~X-4 was observed with the same instrument 
(ACIS back-illuminated chip) at a luminosity within a factor of a few 
of XTE~J1550--564 \citep{rutledge01}.  A soft component and a power-law 
are necessary to fit the Cen~X-4 spectrum in contrast to XTE~J1550--564 
where the spectrum only requires one component.  It is very likely that the 
Cen~X-4 soft component is thermal emission from the surface of the neutron 
star, and it can be modeled as a blackbody with a temperature of 0.175~keV.
Using fit parameters from \cite{rutledge01}, the blackbody component
contributes 62\% of the total 0.5-7~keV unabsorbed flux for Cen~X-4.  
Although a two component model does not improve the fit for XTE~J1550--564, 
we fitted the spectrum with a blackbody plus power-law model to make a more 
direct comparison with Cen~X-4.  The blackbody parameters are not 
well-constrained for XTE~J1550--564, and we fixed the temperature to 
0.175~keV to calculate an upper limit on the blackbody flux.  The 90\% 
confidence upper limit on the ratio of the blackbody flux to the total 
0.5-7~keV unabsorbed flux is 32\%, which is significantly lower that the 
measured value for Cen~X-4.  Finally, we note that the Cen~X-4 power-law 
component has a photon index of $1.2^{+0.4}_{-0.5}$, which is considerably 
harder than the power-law observed for XTE~J1550--564.  We conclude that the 
Cen~X-4 and XTE~J1550--564 spectra are significantly different and speculate 
that this is related to the fact that one contains a neutron star while the 
other contains a black hole.

\subsection{High Energy Cutoff}

Our results indicate that the high energy cutoff for XTE~J1550--564 is stronger 
in the HEXTE band during the state transition (observations 2, 3 and 4) than 
after the source reaches the hard state (observations 5-12).  It may be possible 
to explain this in the context of the standard picture that the hard X-ray 
emission is due to inverse Componization of soft photons by thermal electrons 
where the measured folding energy ($E_{fold}$) is close to the electron 
temperature.  In this picture, the lower value of $E_{fold}$ during the transition 
indicates a lower electron temperature during the transition than in the hard state.  
The temperature change could be related to the observed drop in the strength of 
the soft component during the transition to the hard state since a drop in soft 
photon cooling would lead to a higher temperature for the Comptonizing 
electrons.  However, the lack of a significant cutoff for observation 1 is a 
problem for this picture since the strongest soft component is seen in this case.
A different mechanism, such as Bulk-motion Comptonization \citep{ct95}, 
may be necessary to explain the hard X-ray emission seen during observation 1.  
The possibility that both of these mechanisms may operate in accreting BHCs 
has been discussed previously by several authors (e.g., Ebisawa, Titarchuk
\& Chakrabarti 1996\nocite{etc96}).

The folding energy we obtain for observations 5-12
($E_{fold} = 462^{+342}_{-241}$~keV, 68\% confidence errors) is high 
compared to the folding energies reported by \cite{grove98} for accreting
BHCs in the hard state, which are between 87 and 132~keV.  Since it is 
thought that the high energy emission in the hard state is produced via thermal 
Comptonization, this raises the question of whether realistic accretion disk 
models can be constructed with high enough electron temperatures to give
the values of $E_{fold}$ we observe for XTE~J1550--564.  Detailed 
comparisons between ADAF predictions and the XTE~J1550--564 spectra 
will be a subject of future work, but a preliminary result is that the ADAF 
model under-predicts the level of high energy emission \citep{et01}.  
Emission mechanisms other than thermal Comptonization may be required to 
explain the high energy emission such as Comptonization from a non-thermal 
electron distribution or possibly synchrotron emission \citep{markoff01}.  Since 
the compact jet is a possible source of high energy emission, correlations between 
hard X-ray emission and radio or IR emission (such as those mentioned in \S 2 of 
this paper) are especially interesting.

\section{Summary and Conclusions}

We report on spectral analysis of {\em RXTE} and {\em Chandra} observations 
made during the decay of the 2000 outburst from the BHC XTE~J1550--564.  
A rapid and approximately exponential drop in flux was observed early in 
the decay with an e-folding time of 6.0~days.  The evolution of the energy
spectrum during this time indicates that the system made a transition to
the hard state with a drop in the flux of the soft component in the 
{\em RXTE} energy band and a hardening of the power-law used to model
the hard component.  The transition occurred near a luminosity of
$9\times 10^{36}$~erg~s$^{-1}$ assuming a source distance of 4~kpc.
The spectral changes, the transition luminosity and the time scale
for the transition are similar to those we have previously observed
for another BHC, 4U~1630--47.  However, we note that the 
XTE~J1550--564 hard component behavior was significantly different
during the decay of the 1998-1999 outburst.

The mean luminosity during the {\em Chandra} observations is within 
a factor of a few of the quiescent luminosity of the neutron star system 
Cen~X-4.  In contrast to Cen~X-4 \citep{rutledge01}, the XTE~J1550--564 
spectrum is well-described by a single power-law component with a photon 
index ($\Gamma$) of $2.30^{+0.41}_{-0.48}$ (90\% confidence).  Since 
we measured a value of 1.6 for $\Gamma$ in the hard state, the {\em Chandra} 
spectrum provides reasonably good evidence for spectral softening at low 
luminosities.  ADAF models predict gradual spectral softening as the 
luminosity drops, but our observations do not allow us to determine if the 
spectral evolution is gradual or sudden.  Future observations to measure the 
spectral evolution at intermediate luminosities are important and may allow us 
to distinguish between ADAF models and magnetic flare models.  The lowest 
luminosity we measure for XTE~J1550--564 with {\em Chandra} is 
$5\times 10^{32}$~erg~s$^{-1}$.  This is probably not the true quiescent 
luminosity, but it represents a useful upper limit on this quantity.

Although a highly significant break is seen in the HEXTE band for 
observation 3 during the state transition, the break becomes weaker once 
the source reaches the hard state.  For XTE~J1550--564, the hard state 
cutoff energy is higher than those found for other BHCs \citep{grove98}, 
and this may indicate the presence of non-thermal emission.  More 
theoretical and observational work is necessary to understand the 
mechanism responsible for the high energy emission.  Observations of 
BHC X-ray transients in their hard state with {\em INTEGRAL} will be 
especially useful.

\acknowledgements

We would like to thank the {\em Chandra} Director Harvey Tananbaum 
for granting Director's Discretionary Time and Jean Swank for 
assistance with {\em RXTE} observations.  We also thank Raj Jain 
for providing optical and IR results prior to publication.  
JAT acknowledges useful discussions with Ann Esin.  This material 
is based upon work supported by the National Aeronautics and Space 
Administration under grant NAG5-10886.  PK acknowledges partial 
support from NASA grant NAG5-7405.



\begin{table}
\caption{{\em RXTE} Observations of XTE~J1550--564 \label{tab:rxte_obs}}
\begin{minipage}{8.5in}
\footnotesize
\begin{tabular}{c|c|c|c|c|c|c} \hline \hline
 & & Integration & & PCA Count & HEXTE Count\\
Observation & MJD\footnote{Modified Julian Date (MJD = JD--2400000.5) at the midpoint of the observation.} & 
	Time (s) & PCUs\footnote{The Proportional Counter Units (other than PCU 0) that were on during
	the observation.} & Rate (cps)\footnote{3-25~keV count rate normalized to the rate for 5 PCUs (all 3 layers) 
	after background subtraction.} & Rate (cps)\footnote{18-150~keV count rate for HEXTE (cluster A only)
	after background subtraction.} & Flux\footnote{Absorbed 3-25~keV flux in units of erg~cm$^{-2}$~s$^{-1}$.}\\ 
	\hline
1 & 51680.391 & 2032 & 2,3,4 & 1992 & 36 & $5.6\times 10^{-9}$\\
2 & 51682.316 & 1712 & 2,3,4 & 1575 & 40 & $4.7\times 10^{-9}$\\
3 & 51683.786 & 2688 & 2,3,4 & 1278 & 40 & $4.0\times 10^{-9}$\\
4 & 51684.769 & 848 & 1,2,3,4 & 1145 & 38 & $3.5\times 10^{-9}$\\
5 & 51686.302 & 1520 & 1,2,3,4 & 955 & 33 & $3.0\times 10^{-9}$\\
6 & 51687.229 & 1408 & 2,3 & 855 & 30 & $2.7\times 10^{-9}$\\
7 & 51688.846 & 1712 & 2,3 & 663 & 24 & $2.1\times 10^{-9}$\\
8 & 51690.807 & 1760 & 2,3 & 508 & 17 & $1.6\times 10^{-9}$\\
9 & 51692.562 & 1728 & 2 & 380 & 13 & $1.2\times 10^{-9}$\\
10 & 51693.411 & 1168 & 2 & 330 & 12 & $1.0\times 10^{-9}$\\
11 & 51695.271 & 1136 & 2,3 & 258 & 9 & $8.1\times 10^{-10}$\\
12 & 51696.483 & 1712 & 2,3,4 & 218 & 8 & $7.1\times 10^{-10}$\\
13 & 51698.948 & 1152 & 2 & 165 & 6 & $5.0\times 10^{-10}$\\
\end{tabular}
\end{minipage}
\end{table}

\begin{table}
\caption{{\em RXTE} Measurement of Power-Law Photon Index ($\Gamma$) \label{tab:spec}}
\begin{minipage}{4.5in}
\footnotesize
\begin{tabular}{c|c|c|c|c|c} \hline \hline
 & & & PCA+HEXTE & PCA+HEXTE & \\
Observation & PCA\footnote{68\% confidence errors are given.} & HEXTE$^{a}$ & (No Cutoff)$^{a}$ & 
	(With Cutoff)$^{a}$ & Consensus\footnote{The measured value for $\Gamma$ 
with PCA and HEXTE after accounting for the high energy cutoffs detected for observations 2, 3 and 4.  
The error given includes the systematic error (see text).}\\ \hline
1 & $2.047\pm 0.012$ & $1.997\pm 0.022$ & $2.053\pm 0.010$ & - & $2.05\pm 0.04$\\
2 & $1.864\pm 0.013$ & $1.849\pm 0.020$ & $1.878\pm 0.011$ & $1.852\pm 0.014$ & $1.85\pm 0.04$\\
3 & $1.717\pm 0.005$ & $1.698\pm 0.013$ & $1.720\pm 0.005$ & $1.714\pm 0.004$ & $1.71\pm 0.04$\\
4 & $1.698\pm 0.006$ & $1.674\pm 0.023$ & $1.700\pm 0.006$ & $1.696\pm 0.005$ & $1.70\pm 0.04$\\
5 & $1.669\pm 0.005$ & $1.580\pm 0.021$ & $1.665\pm 0.005$ & - & $1.67\pm 0.04$\\
6 & $1.640\pm 0.008$ & $1.597\pm 0.025$ & $1.637\pm 0.007$ & - & $1.64\pm 0.04$\\
7 & $1.606\pm 0.007$ & $1.578\pm 0.023$ & $1.606\pm 0.007$ & - & $1.61\pm 0.04$\\
8 & $1.609\pm 0.008$ & $1.601\pm 0.027$ & $1.609\pm 0.007$ & - & $1.61\pm 0.04$\\
9 & $1.591\pm 0.013$ & $1.634\pm 0.041$ & $1.598\pm 0.012$ & - & $1.60\pm 0.04$\\
10 & $1.599\pm 0.018$ & $1.469\pm 0.053$ & $1.581\pm 0.016$ & - & $1.58\pm 0.05$\\
11 & $1.640\pm 0.018$ & $1.559\pm 0.070$ & $1.632\pm 0.017$ & - & $1.63\pm 0.05$\\
12 & $1.620\pm 0.012$ & $1.627\pm 0.065$ & $1.620\pm 0.012$ & - & $1.62\pm 0.05$\\
13 & $1.718\pm 0.021$ & $1.629\pm 0.171$ & $1.726\pm 0.033$ & - & $1.73\pm 0.06$\\
\end{tabular}
\end{minipage}
\end{table}

\begin{table}
\caption{Break Parameters for the {\em RXTE} Spectra\label{tab:cutoff}}
\begin{minipage}{\linewidth}
\footnotesize
\begin{tabular}{c|c|c} \hline \hline
Observation & $E_{cut}$ (keV)\footnote{68\% confidence errors are given.} & 
	$E_{fold}$ (keV)$^{a}$\\ \hline
2 & $40^{+8}_{-7}$ & $338^{+91}_{-83}$\\
3 & $72^{+8}_{-9}$ & $194^{+71}_{-48}$\\
4 & $73^{+10}_{-18}$ & $175^{+110}_{-57}$\\
5-12 & $78^{+23}_{-11}$ & $462^{+342}_{-241}$\\
\end{tabular}
\end{minipage}
\end{table}


\begin{thebibliography}{}
\setlength{\itemsep}{0em}
\setlength{\parsep}{0pt}
\setlength{\baselineskip}{0pt}
\renewcommand{\baselinestretch}{0.5}
\footnotesize

\bibitem[\protect\astroncite{{Asai} et~al.}{1998}]{asai98}
{Asai}, K., {Dotani}, T., {Hoshi}, R., {Tanaka}, Y., {Robinson}, C.~R., \&
  {Terada}, K.,  1998, \pasj, 50, 611

\bibitem[\protect\astroncite{{Balbus} \& {Hawley}}{1991}]{bh91}
{Balbus}, S.~A., \& {Hawley}, J.~F.,  1991, \apj, 376, 214

\bibitem[\protect\astroncite{{Ball}, {Narayan} \& {Quataert}}{2001}]{bnq01}
{Ball}, G.~H., {Narayan}, R., \& {Quataert}, E.,  2001, \apj, 552, 221

\bibitem[\protect\astroncite{{Bradt}, {Rothschild} \& {Swank}}{1993}]{brs93}
{Bradt}, H.~V., {Rothschild}, R.~E., \& {Swank}, J.~H.,  1993, A\&AS, 97, 355

\bibitem[\protect\astroncite{{Campana}, {Parmar} \& {Stella}}{2001}]{cps01}
{Campana}, S., {Parmar}, A.~N., \& {Stella}, L.,  2001, \aap, 372, 241

\bibitem[\protect\astroncite{{Campbell-Wilson} et~al.}{1998}]{campbell98}
{Campbell-Wilson}, D., {McIntyre}, V., {Hunstead}, R., {Green}, A., {Wilson},
  R.~B., \& {Wilson}, C.~A.,  1998, IAU~Circular, 7010

\bibitem[\protect\astroncite{{Cash}}{1979}]{cash79}
{Cash}, W.,  1979, \apj, 228, 939

\bibitem[\protect\astroncite{{Chakrabarti} \& {Titarchuk}}{1995}]{ct95}
{Chakrabarti}, S., \& {Titarchuk}, L.~G.,  1995, \apj, 455, 623

\bibitem[\protect\astroncite{{Chen}, {Shrader} \& {Livio}}{1997}]{csl97}
{Chen}, W., {Shrader}, C.~R., \& {Livio}, M.,  1997, \apj, 491, 312

\bibitem[\protect\astroncite{{Corbel} et~al.}{2001}]{corbel01}
{Corbel}, S., et~al., 2001, \apj, 554, 43

\bibitem[\protect\astroncite{Cui et~al.}{1999}]{cui99}
Cui, W., Zhang, S.~N., Chen, W., \& Morgan, E.~H.,  1999, \apj, 512, L43

\bibitem[\protect\astroncite{{di Matteo}, {Celotti} \& {Fabian}}{1999}]{dcf99}
{di Matteo}, T., {Celotti}, A., \& {Fabian}, A.~C.,  1999, \mnras, 304, 809

\bibitem[\protect\astroncite{{Ebisawa} et~al.}{1994}]{ebisawa94}
{Ebisawa}, K., et~al., 1994, \pasj, 46, 375

\bibitem[\protect\astroncite{{Ebisawa}, {Titarchuk} \&
  {Chakrabarti}}{1996}]{etc96}
{Ebisawa}, K., {Titarchuk}, L., \& {Chakrabarti}, S.~K.,  1996, \pasj, 48, 59

\bibitem[\protect\astroncite{{Esin}, {McClintock} \& {Narayan}}{1997}]{emn97}
{Esin}, A.~A., {McClintock}, J.~E., \& {Narayan}, R.,  1997, ApJ, 489, 865

\bibitem[\protect\astroncite{{Esin} \& {Tomsick}}{2001}]{et01}
{Esin}, A.~A., \& {Tomsick}, J.~A.,  2001,
\newblock in American Astronomical Society Meeting, Vol. 198,  1001

\bibitem[\protect\astroncite{{Galeev}, {Rosner} \& {Vaiana}}{1979}]{grv79}
{Galeev}, A.~A., {Rosner}, R., \& {Vaiana}, G.~S.,  1979, \apj, 229, 318

\bibitem[\protect\astroncite{{Gammie} \& {Menou}}{1998}]{gm98}
{Gammie}, C.~F., \& {Menou}, K.,  1998, \apjl, 492, L75

\bibitem[\protect\astroncite{{Garcia} et~al.}{2001}]{garcia01}
{Garcia}, M.~R., {McClintock}, J.~E., {Narayan}, R., {Callanan}, P., {Barret},
  D., \& {Murray}, S.~S.,  2001, \apjl, 553, L47

\bibitem[\protect\astroncite{{Grove} et~al.}{1998}]{grove98}
{Grove}, J.~E., {Johnson}, W.~N., {Kroeger}, R.~A., {McNaron-Brown}, K.,
  {Skibo}, J.~G., \& {Phlips}, B.~F.,  1998, \apj, 500, 899

\bibitem[\protect\astroncite{{Hannikainen} et~al.}{2001}]{hannikainen01}
{Hannikainen}, D., et~al., 2001,
\newblock in Proceedings of the 4th INTEGRAL Workshop (Alicante 2000), to be
  published in ESA-SP (2001),  2070

\bibitem[\protect\astroncite{{Homan}, {Wijnands} \& {van der
  Klis}}{1999}]{homan99}
{Homan}, J., {Wijnands}, R., \& {van der Klis}, M.,  1999, \iaucirc, 7121

\bibitem[\protect\astroncite{{Homan} et~al.}{2001}]{homan01}
{Homan}, J., {Wijnands}, R., {van der Klis}, M., {Belloni}, T., {van Paradijs},
  J., {Klein-Wolt}, M., {Fender}, R., \& {M{\'e}ndez}, M.,  2001, ApJS, 132,
  377

\bibitem[\protect\astroncite{{Jain}, {Bailyn} \& {Tomsick}}{2001}]{jbt01}
{Jain}, R., {Bailyn}, C., \& {Tomsick}, J.,  2001, \iaucirc, 7575

\bibitem[\protect\astroncite{{Jain} et~al.}{2001}]{jain01}
{Jain}, R.~K., {Bailyn}, C.~D., {Orosz}, J.~A., {McClintock}, J.~E., \&
  {Remillard}, R.~A.,  2001, \apjl, 554, L181

\bibitem[\protect\astroncite{{Jain} et~al.}{1999}]{jain99}
{Jain}, R.~K., {Bailyn}, C.~D., {Orosz}, J.~A., {Remillard}, R.~A., \&
  {McClintock}, J.~E.,  1999, ApJ, 517, L131

\bibitem[\protect\astroncite{Kalemci et~al.}{2001}]{kalemci01}
Kalemci, E., Tomsick, J., Rothschild, R., Pottschmidt, K., \& Kaaret, P.,
  2001, ApJ, submitted, paper II

\bibitem[\protect\astroncite{{Kaptein}, {in 't Zand} \&
  {Heise}}{2001}]{kaptein01}
{Kaptein}, R., {in 't Zand}, J. J.~M., \& {Heise}, J.,  2001, \iaucirc, 7588

\bibitem[\protect\astroncite{{Knight}}{1982}]{knight82}
{Knight}, F.~K.,  1982, \apj, 260, 538

\bibitem[\protect\astroncite{{Kong} et~al.}{2000}]{kong00}
{Kong}, A. K.~H., {Kuulkers}, E., {Charles}, P.~A., \& {Homer}, L.,  2000,
  \mnras, 312, L49

\bibitem[\protect\astroncite{{Kotani} et~al.}{2000}]{kotani00}
{Kotani}, T., et~al., 2000, \apjl, 543, L133

\bibitem[\protect\astroncite{{Lightman} \& {White}}{1988}]{lw88}
{Lightman}, A.~P., \& {White}, T.~R.,  1988, \apj, 335, 57

\bibitem[\protect\astroncite{{Makishima} et~al.}{1986}]{makishima86}
{Makishima}, K., {Maejima}, Y., {Mitsuda}, K., {Bradt}, H.~V., {Remillard},
  R.~A., {Tuohy}, I.~R., {Hoshi}, R., \& {Nakagawa}, M.,  1986, \apj, 308, 635

\bibitem[\protect\astroncite{{Markoff}, {Falcke} \& {Fender}}{2001}]{markoff01}
{Markoff}, S., {Falcke}, H., \& {Fender}, R.,  2001, \aap, 372, L25

\bibitem[\protect\astroncite{{Marshall}, {Takeshima} \& {in 't
  Zand}}{2000}]{marshall00}
{Marshall}, F.~E., {Takeshima}, T., \& {in 't Zand}, J.,  2000, \iaucirc, 7363

\bibitem[\protect\astroncite{{Massaro} et~al.}{2000}]{massaro00}
{Massaro}, E., {Cusumano}, G., {Litterio}, M., \& {Mineo}, T.,  2000, \aap,
  361, 695

\bibitem[\protect\astroncite{Miller et~al.}{2001a}]{miller01a}
Miller, J., et~al., 2001a, \mnras, submitted, astro-ph/0103215

\bibitem[\protect\astroncite{Miller et~al.}{2001b}]{miller01b}
Miller, J., et~al., 2001b, \apjl, submitted, astro-ph/0105371

\bibitem[\protect\astroncite{{Narayan}, {Barret} \& {McClintock}}{1997}]{nbm97}
{Narayan}, R., {Barret}, D., \& {McClintock}, J.~E.,  1997, \apj, 482, 448

\bibitem[\protect\astroncite{{Narayan}, {Garcia} \& {McClintock}}{1997}]{ngm97}
{Narayan}, R., {Garcia}, M.~R., \& {McClintock}, J.~E.,  1997, \apjl, 478, L79

\bibitem[\protect\astroncite{{Narayan}, {Igumenshchev} \&
  {Abramowicz}}{2000}]{nia00}
{Narayan}, R., {Igumenshchev}, I.~V., \& {Abramowicz}, M.~A.,  2000, \apj, 539,
  798

\bibitem[\protect\astroncite{{Orosz}, {Bailyn} \& {Jain}}{1998}]{obj98}
{Orosz}, J., {Bailyn}, C., \& {Jain}, R.,  1998, IAU~Circular, 7009

\bibitem[\protect\astroncite{{Orosz} et~al.}{2001}]{orosz01}
{Orosz}, J.~A., {van der Klis}, M., {McClintock}, J.~E., {Jain}, R.~K.,
  {Bailyn}, C.~D., \& {Remillard}, R.~A.,  2001, The Astronomer's Telegram,
  \#70, 70

\bibitem[\protect\astroncite{{Predehl} \& {Schmitt}}{1995}]{ps95}
{Predehl}, P., \& {Schmitt}, J. H. M.~M.,  1995, \aap, 293, 889

\bibitem[\protect\astroncite{{Remillard} et~al.}{1999}]{remillard99}
{Remillard}, R.~A., {McClintock}, J.~E., {Sobczak}, G.~J., {Bailyn}, C.~D.,
  {Orosz}, J.~A., {Morgan}, E.~H., \& {Levine}, A.~M.,  1999, ApJ, 517, L127

\bibitem[\protect\astroncite{Remillard et~al.}{2001}]{remillard01}
Remillard, R.~A., Sobczak, G.~J., Muno, M.~P., \& McClintock, J.~E.,  2001,
  ApJ, submitted, astro-ph/0105508

\bibitem[\protect\astroncite{Rothschild et~al.}{1998}]{rothschild98}
Rothschild, R.~E., et~al., 1998, ApJ, 496, 538

\bibitem[\protect\astroncite{{Rutledge} et~al.}{2001}]{rutledge01}
{Rutledge}, R.~E., {Bildsten}, L., {Brown}, E.~F., {Pavlov}, G.~G., \&
  {Zavlin}, V.~E.,  2001, \apj, 551, 921

\bibitem[\protect\astroncite{{S{\'a}nchez-Fern{\'a}ndez}
  et~al.}{1999}]{sanchez99}
{S{\'a}nchez-Fern{\'a}ndez}, C., et~al., 1999, \aap, 348, L9

\bibitem[\protect\astroncite{{Smith}}{1998}]{smith98}
{Smith}, D.~A.,  1998, IAU~Circular, 7008

\bibitem[\protect\astroncite{{Smith} et~al.}{2000}]{smith00}
{Smith}, D.~A., {Levine}, A.~M., {Remillard}, R., {Fox}, D., {Schaefer}, R., \&
  {RXTE/ASM Team} 2000, IAU~Circular, 7399

\bibitem[\protect\astroncite{{Sobczak} et~al.}{2000}]{sobczak00}
{Sobczak}, G.~J., {McClintock}, J.~E., {Remillard}, R.~A., {Cui}, W., {Levine},
  A.~M., {Morgan}, E.~H., {Orosz}, J.~A., \& {Bailyn}, C.~D.,  2000, \apj, 544,
  993

\bibitem[\protect\astroncite{{Sobczak} et~al.}{1999}]{sobczak99}
{Sobczak}, G.~J., {McClintock}, J.~E., {Remillard}, R.~A., {Levine}, A.~M.,
  {Morgan}, E.~H., {Bailyn}, C.~D., \& {Orosz}, J.~A.,  1999, ApJ, 517, L121

\bibitem[\protect\astroncite{{{Tanaka}, Y. and {Lewin}, W. H. G.}}{1995}]{tl95}
{{Tanaka}, Y. and {Lewin}, W. H. G.},  1995,
\newblock {X}-ray {B}inaries,
\newblock  Cambridge: Cambridge {U}. {P}ress

\bibitem[\protect\astroncite{{Titarchuk}}{1994}]{titarchuk94}
{Titarchuk}, L.,  1994, \apj, 434, 570

\bibitem[\protect\astroncite{{Tomsick} \& {Kaaret}}{2000}]{tk00}
{Tomsick}, J.~A., \& {Kaaret}, P.,  2000, ApJ, 537, 448

\bibitem[\protect\astroncite{{Tomsick} et~al.}{2001}]{tomsick01}
{Tomsick}, J.~A., {Smith}, E., {Swank}, J., {Wijnands}, R., \& {Homan}, J.,
  2001, IAU~Circular, 7575

\bibitem[\protect\astroncite{{Valinia} \& {Marshall}}{1998}]{vm98}
{Valinia}, A., \& {Marshall}, F.~E.,  1998, \apj, 505, 134

\bibitem[\protect\astroncite{{{van der Klis}, M.}}{1995}]{vdk95}
{{van der Klis}, M.},  1995,
\newblock {X}-ray {B}inaries,
\newblock  Cambridge: Cambridge {U}. {P}ress

\bibitem[\protect\astroncite{{Wilson} \& {Done}}{2001}]{wd01}
{Wilson}, C.~D., \& {Done}, C.,  2001, \mnras, 325, 167

\end{thebibliography}
\end{document}